\documentclass[final]{ias2}

\usepackage{graphicx} 
\usepackage{multirow}
\usepackage{array} 

\usepackage{hyperref} 

\begin{document}

\markboth{Dynamo in Protostar}{Verma, et. al.}

\title{Dynamo in Protostar}

\author[1]{Mahendra K. Verma} 
\email{mkv@iitk.ac.in}
\author[2]{Bidya Binay Karak} 
\email{bidya\_karak@physics.iisc.ernet.in}
\author[1]{Rohit Kumar} 
\email{rohitkr@iitk.ac.in}
\address[1]{Department of Physics, Indian Institute of Technology, Kanpur 208016, India}
\address[2]{Department of Physics, Indian Institute of Science, Bangalore 560012, India}

\begin{abstract}
In this paper, we estimate the magnetic Reynolds number of a typical protostar  before and after deuterium burning, and claim for the existence of dynamo process in both the phases, because the magnetic Reynolds number of the protostar far exceeds the critical magnetic Reynolds number for dynamo action. Using the equipartition of kinetic and magnetic energies, we estimate the steady-state magnetic field of the protostar to be of the order of kilo-gauss, which is in good agreement with observations. 
\end{abstract}


\pacs{Appropriate pacs here}
 
\maketitle

\section{Introduction}
Magnetic field is omnipresent in the universe.  It is found in many stars, planets,  galaxies, interstellar medium, etc., and it is believed to induce inflow of matter or accretion in astrophysical disc and immersed cloud, collimated jets and outflows, magnetic breaking, and protostellar winds.    The magnetic field plays a crucial role in all stages of the  evolution  of a protostar.  In this paper we will address an important question related to protostars: when does the magnetic field appears in protostars, and what is the magnitude of the magnetic field, if present?

Presence of X-ray from the protostar regions suggests a strong magnetic field in its central region \cite{Montmerle, Hamaguchi}.   Also, the detection of polarized synchrotron emission arising from protostar jets \cite{Carrasco-Gonzalez} and the measurements from Zeeman broadening of photospheric lines \cite{Johns-Krull, Guenther, Yang, Bouvier} provide a strong support for the existence of the magnetic field in the protostars.   Several questions on protostars are: whether the magnetic field arises from the molecular cloud of the protostar formation region, or due to self induction or dynamo mechanism? How is the magnetic field sustains itself and compensates its removal due to the outflows, or its dissipation as ohmic heating? Does the dynamo mechanism, if any, stops after the formation of the protostars?

Machida et al.~\cite{Machida} and Sur et al.~\cite{Sur} argue that the dynamo mechanism can amplify the initial magnetic field of the star forming regions. On the other hand, Tan and Blackman \cite{Tan} suggest that dynamo amplification of the primordial magnetic field is possible in protostellar disc, and they propose that the induced magnetic field is helical. In these work, the generation mechanism for the magnetic field involves several mechanisms, e.g., convection, rotation, etc. 
However, there is no substantive statement in the literature on the necessity of dynamo process in protostars.  In the present paper we claim that the protostars are magnetic because its magnetic Reynolds number to trigger the dynamo is far above the critical magnetic Reynolds number observed in laboratory experiments and numerical simulations.  The equipartition of kinetic and magnetic energies yields magnetic field of the protostar to be of the order of kilo-gauss, which is in good agreement to those observed in nature.  We present these arguments in the next two sections of the paper, and conclude in the last section.

\section{Critical values of magnetic Reynolds number for dynamo}

The fluid velocity $\mathbf u$ and the magnetic field $\mathbf B$ in a dynamo mechanism are governed by  the magnetohydrodynamic (MHD) equations:
\begin{eqnarray}
\rho\{ \partial_{t}\mathbf{u}+ (\mathbf{u} \cdot \nabla) \mathbf{u} \} & = &
-\nabla p+ (\mathbf{J} \times \mathbf{B} ) + \nu\nabla^{2}\mathbf{u}+\mathbf{F}, \label{eq:MHD_vel}\\
\partial_{t}\mathbf{B} & = &
\nabla \times ( \mathbf{u} \times \mathbf {B} )+\eta\nabla^{2}\mathbf{B}, \\
\nabla \cdot \mathbf{B} & = & 0,  \label{eq:div_B_0}
\end{eqnarray}
where $\mathbf u$, $\mathbf J$, $\mathbf F$, $p$, $\nu$, and $\eta$ represent the velocity field, current density, external forcing, hydrodynamic pressure, kinematic viscosity, and magnetic diffusivity, respectively.  Note that ${\bf J} = c (\nabla \times {\bf B}) / 4 \pi$.  The density of the fluid is governed by the continuity equation, though some computations assume it to be a constant~\cite{Glatzmaier}.  Note that only the velocity field is being forced in the above equation.  

Dynamo mechanism is said to occur when the magnetic energy reaches a finite value asymptotically, that is as $t\rightarrow \infty$.  The magnetic field gets energy from the velocity field through the $\nabla \times ( \mathbf{u} \times \mathbf {B} )$ term, and loses energy via Joule dissipation.  For magnetic energy to grow and reach a steady state, a comparison of these two terms suggests that
\begin{equation}
(\nabla \times ( \mathbf{u} \times \mathbf {B} ))  > \eta\nabla^{2}\mathbf{B} 
\end{equation}
or, the magnetic Reynolds number
\begin{equation}
\mathrm{Rm} = \frac{U L}{\eta} > 1
\end{equation}
The minimum magnetic Reynolds number, called the {\em critical magnetic Reynolds number} $\mathrm{Rm}_c$, could be larger than one due to geometrical consideration or other factors like magnetic Prandtl number ($\mathrm{Pm}$, defined as $\nu/\eta$), rotation, etc.  Yet, we can argue that $\mathrm{Rm}_c$  is of the order of unity for all dynamo processes; this limit has been consistently observed in a number of laboratory experiments and numerical simulations.  

Among laboratory experiments, Gailities (Riga experiment) {\it et al.}~\cite{Gailitis}, Stieglitz and M\"{u}ller (Karlsruhe experiment)~\cite{Stieglitz}, and Monchaux {\em et al.} (VKS experiment)~\cite{monchaux} observed self-generated magnetic fields in their experimental setups.  Liquid sodium, whose magnetic Prandtl number is approximately $10^{-5}$, was used as the operating fluid in the aforementioned experiments.  The critical magnetic Reynolds number for the above set of experiments was greater than 10; specifically, the $\mathrm{Rm}_c \approx 30 $ for that Monchaux {\em et al.}'s experiment, commonly referred to as VKS experiment~\cite{monchaux}.

 A large number of high-resolution simulations have been performed to study the dynamo transition.  Most researchers used pseudo-spectral method to solve the MHD equations for various forcing in a box geometry.  The range of magnetic Prandtl number used so far is from $10^{-2}$ to $10^{2}$.  Note that numerical simulations of very large $\mathrm{Pm}$ or very small $\mathrm{Pm}$ dynamo are difficult due to their requirements of high resolutions.   Notably, Schekochihin {\it et al.}~\cite{schekochihin_1,schekochihin_2}, Ponty {\it et al.}~\cite{Ponty}, and Iskakov {\em et al.}~\cite{Iskakov} simulated dynamo and studied variation of $\mathrm{Rm}_c$  as a function of magnetic Prandtl number.  They observed dynamo for both small and large Prandtl numbers, with the range of $\mathrm{Rm}_c$ between 10 and 500.  The $\mathrm{Rm}_c$  for lower Prandtl number tends to higher in most of the simulations.   Recently, Yadav et al.~\cite{rakesh_1,rakesh_2} studied Taylor-Green dynamo and  observed the $\mathrm{Rm}_c$ to be around 10.     In Fig.~1 we plot,  $\mathrm{Rm}_c$ for the aforementioned experiments and numerical simulations.  We can conclude from these results and the aforementioned phenomenological arguments relating the nonlinear term to the dissipative term that $\mathrm{Rm}_c$ varies with geometry, forcing, and Prandtl number, but it remains bounded between 1 and 500.  Hence, if the protostars have $\mathrm{Rm} > \mathrm{Rm}_c$, then dynamo must be active in the protostar.  This is what we would estimate in the next section.   

\begin{figure}[h]
\begin{center}
\includegraphics[scale=0.5]{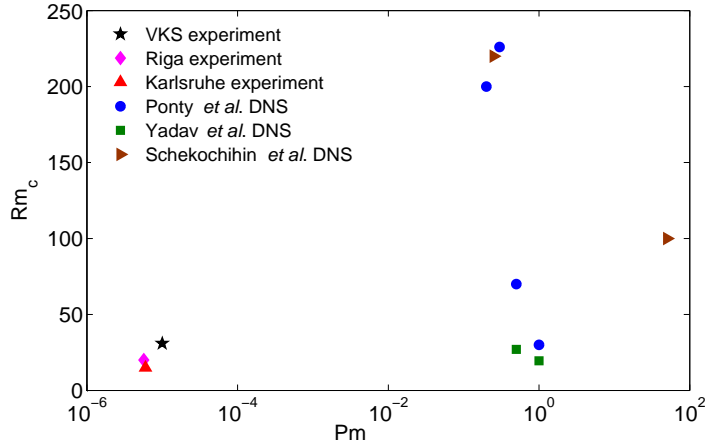}
\caption{Plot of critical magnetic Reynolds number ($\mathrm{Rm}_c$) vs. magnetic Prandtl number ($\mathrm{Pm}$).  The plot exhibits data of VKS experiment~\cite{monchaux}, Riga experiment~\cite{Gailitis}, Karlsruhe experiment~\cite{Stieglitz},  and numerical data of Ponty {\em et al.}~DNS~\cite{Ponty}, Yadav {\it et al.}~DNS~\cite{rakesh_1, rakesh_2}, and Schekochihin {\it et al.}~DNS~\cite{schekochihin_1, schekochihin_2}.}
\end{center}
\label{fig:rmc}
\end{figure}

\section{Magnetic  Reynolds number and dynamo action in the protostar}
In this section we will discuss the dynamo process in protostar before and after deuterium burning.     First, we focus on a typical protostar before deuterium burning.  Temperature of the protostar  at this stage may be around  $10^{4}$~K, its radius around 100 solar-radii, and its mass around two solar mass. The kinematic viscosity and the resistivity of the gas can be estimated using Spitzer formula (see e.g., Schekochihin {\it et al.} \cite{schekochihin05} or Choudhuri \cite{choudhuri}) that provides
\begin{eqnarray}
\nu & \sim & 2.21\times10^{-15}T^{5/2}/(4\rho),\\
\eta & \sim & 4\pi^{3/2}m_{e}^{1/2}e^{2}c^{2}/(2\times(2k_{B}T)^{3/2}\times0.6),
\end{eqnarray}
where $\rho$ is the density of the medium, $T$ is the temperature in Kelvin, $k_{B}$ is the Boltzmann's constant, $m_{e}$ and $e$ are the mass and charge of an electron, and $c$ is the speed of light. Note that the above formulas are in CGS units.  We substitute the aforementioned  parameters, $T = 10^{4}$~K and the average density $\rho \approx 3 \times 10^{-6}~ \mathrm{gm/cm^3}$, in the above formulas.   Also, we take the large-length scale $L$ of the protostar to be $10^{10} $ cm, and the large-scale velocity $U$ to be one-tenth of the sound speed. Consequently,
\begin{eqnarray}
\nu & \approx & 2~ \mathrm{cm^2/s} ,\\
\eta & \approx & 3 \times 10^7~ \mathrm{cm^2/s}  , \\
U & \approx & 10^5~ \mathrm{cm/s}, \\
\mathrm{Rm} &  \approx & 5 \times 10^7, \\
\mathrm{Pm} & \approx & 10^{-7}.
\end{eqnarray}
The magnetic Reynolds number $\mathrm{Rm}$ is much greater than the estimated critical magnetic Reynolds number ($10-500$) using the laboratory experimental and numerical data. Hence we expect dynamo to be active in typical  protostars before the deuterium burning stage.  The  background galactic magnetic field or the magnetic field amplified during the formation of protostars could act as the seed magnetic field for this process. 

The saturated or the steady-state magnetic field of the protostar can be estimated easily using the equipartition of kinetic and magnetic energy, i. e., $B^2/4\pi \sim \rho U^2$, which yields the average magnetic field at this stage to be around 700 Gauss.  Note that approximate equipartition of kinetic and magnetic energy in the steady state has been observed in large number of systems (e.g., solar wind) and numerical simulations, hence it is a robust assumption for estimating the steady-state magnetic field.  Numerical simulations also reveal that saturation of the magnetic field takes around 10 eddy turnover time.  Hence we expect the time taken for the magnetic field of the protostar to reach a steady state would be approximately $10 L/U$, which is approximately 12 terrestrial days.

Now let us study the possibility of dynamo process in the protostar after deuterium burning. At this stage, the temperature of a typical protostar is approximately $10^{6}$~K, and its radius around two solar-radius.   Here, we take the large-length scale $L$ to be one tenth of the solar radius, and time period of the large eddies to be the rotation time-period of the protostar, which is approximately 20 terrestrial days.  Hence,
\begin{eqnarray}
\nu & \approx & 2~ \mathrm{cm^2/s} ,\\
\eta & \approx & 3 \times 10^4~ \mathrm{cm^2/s}  , \\
U & \approx & 10^4~ \mathrm{cm/s}, \\
\mathrm{Rm} &  \approx & 6 \times 10^9, \\
\mathrm{Pm} & \approx & 6 \times 10^{-5}.
\end{eqnarray}
Again, since $\mathrm{Rm} > \mathrm{Rm}_c$, we expect dynamo to be active in the protostar after the deuterium burning phase.   The magnetic field generated in the pre deuterium-burning stage would act as the seed magnetic field.  Equipartition of the kinetic and magnetic energy yields the average magnetic field to be of the order of $2\times 10^4$ Gauss, and it would take around 200 terrestrial days for the magnetic field to reach saturation.   Note that our simple estimate of the protostar magnetic field is in general agreement with many observational measurements of Zeeman broadening~\cite{Johns-Krull, Guenther, Yang, Bouvier}, which reveal the existence of magnetic field of kilogauss strength. The high field thus generated could produce X-ray activity, jets, outflows, and other magnetic effects.  We remark that the above estimates are in good agreement with the observed parameters of the Sun.

\section{Conclusions}
In this paper, we estimate the magnetic Reynolds number $\mathrm{Rm}$ of a typical protostar before and after deuterium burring.  We show that the $\mathrm{Rm}$ in both the phases is far greater than the critical magnetic Reynolds number $\mathrm{Rm}_c$, which could be between 10 and 500, an estimate based on dimensional analysis, laboratory experiments, and numerical simulations.  Thus, we claim that the dynamo mechanism is active in protostar in both the phases.  We also estimate the steady-state magnetic field of the protostar in these two phases using the equipartition of kinetic and magnetic energy, and find them to be of the order of a kilo-gauss, which are in very good agreement with  several astronomical observations~\cite{Johns-Krull, Guenther, Yang, Bouvier}.  

Our arguments on protostar dynamo is quite robust, and it is independent of mechanism invoked for the process, e.g., convection, rotation, $\alpha-\omega$, etc.  It must be however kept in mind that the actual magnetic field of the star would depend on the details of the dynamo process.  Our arguments  provide an argument in favour of the dynamo mechanism, as well as an estimate of the strength of the magnetic field, in protostar.

\section*{Acknowledgments}
We are grateful to Own Mathews, Prateek Sharma, Dinshaw Balsara, Banibrata Mukhopadhyay, and Sagar Chakraborty for valuable suggestions and comments at various stages.

\bibliographystyle{pramana}
\bibliography{references}
%

\bibliographystyle{pramana}


\end{document}